\documentclass[english,aps,pre,twocolumn]{revtex4-1}
\usepackage[T1]{fontenc}
\usepackage[latin9]{inputenc}
\usepackage{amstext}
\usepackage{amssymb}
\usepackage{graphicx}

\makeatletter

\providecommand{\tabularnewline}{\\}

 \@ifundefined{textcolor}{}
 {%
   \definecolor{BLACK}{gray}{0}
   \definecolor{WHITE}{gray}{1}
   \definecolor{RED}{rgb}{1,0,0}
   \definecolor{GREEN}{rgb}{0,1,0}
   \definecolor{BLUE}{rgb}{0,0,1}
   \definecolor{CYAN}{cmyk}{1,0,0,0}
   \definecolor{MAGENTA}{cmyk}{0,1,0,0}
   \definecolor{YELLOW}{cmyk}{0,0,1,0}
 }

\makeatother

\usepackage{babel}
\begin{document}

\title{Bending rigidity and higher order curvature terms for the hard-sphere
fluid near a curved wall}

\author{Ignacio Urrutia$^{*\dag}$}
\begin{abstract}
In this work I derive analytic expressions for the curvature dependent
fluid-substrate surface tension of a hard sphere fluid on a hard curved
wall. In a first step, the curvature thermodynamic properties are
found as truncated power series in the activity in terms of the known
second and third order cluster integral of the hard-sphere fluid near
spherical and cylindrical walls. These results are then expressed
as packing fraction power series and transformed to different reference
regions which is equivalent to consider different positions of the
dividing surface. All the obtained series expansions are rigorous
results. Based on the truncated series it is shown that the bending
rigidity of the system is non-null and that higher order terms in
the curvature also exist. In a second step, approximate analytic expressions
for the surface tension, the Tolman length, the bending rigidity and
the Gaussian rigidity as functions of the packing fraction are found
by considering the known terms of the series expansion complemented
with a simple fitting approach. It is found that the obtained formulas
accurately describe the curvature thermodynamic properties of the
system, further, they are more accurate than any of the previously
published expressions.
\end{abstract}

\affiliation{$^{*}$Departamento de Física de la Materia Condensada, Centro Atómico
Constituyentes, CNEA, Av.Gral.~Paz 1499, 1650 Pcia.~de Buenos Aires,
Argentina}

\affiliation{$^{\dag}$ CONICET.}

\maketitle

\section{Introduction}

The relation between the thermodynamic properties of a confined fluid
and the shape of the vessel where it is confined has become a topic
of current interest. Recently, the inhomogeneous fluid cluster expansion
was revisited and applied to the hard sphere (HS) fluid near a hard
planar-wall \cite{Yang_2013}. Moreover, the HS system near spherical
and cylindrical hard walls was studied focusing on the analysis of
the curvature dependent fluid-wall surface tension using both, molecular
dynamics and the density functional theory \cite{Laird_2010,Laird_2012,Blokhuis_2013}.

The studies of the \emph{local} dependence of the surface tension
or surface free-energy on the curvature (for membranes, for two fluid-phases
and also for fluid/wall, systems) rely in one of the two following
different expressions. The first one, derived by Helfrich is \cite{Helfrich_1973}
\begin{equation}
\gamma\left(J,K\right)=\gamma-\delta\gamma\, J+\frac{k}{2}J^{2}+\bar{k}K+\ldots\:,\label{eq:gamJK}
\end{equation}
where the truncation to order $K$ is frequently assumed. The second
one, proposed by König \emph{et al.} and based on the Hadwiger Theorem
\cite{Hadwiger1957} is 
\begin{equation}
\gamma\left(J,K\right)=\gamma-\delta\gamma\, J+\bar{k}K\:.\label{eq:gamHadwiger}
\end{equation}
In Eqs. (\ref{eq:gamJK}, \ref{eq:gamHadwiger}) $J=R_{1}^{-1}+R_{2}^{-1}$
is the total curvature, $K=R_{1}^{-1}R_{2}^{-1}$ is the Gaussian
curvature, and $R_{1}$, $R_{2}$ are the \emph{local} principal radius
of the surface. The fluid-wall surface tension (or surface free-energy)
for the case of a planar wall is indicated by $\gamma$, while the
curvature coefficients are: $\delta$ the Tolman's length, $k$ the
bending rigidity and $\bar{k}$ the Gaussian-curvature rigidity. Both
expressions were successfully utilized to describe, on approximate
grounds, the properties of interfaces. Furthermore, it was conjectured
that Eq. (\ref{eq:gamHadwiger}) is complete for the HS fluid in contact
with hard curved walls \cite{Bryk_2003a,Konig_2004,Konig_2005}.

The analysis of the curvature dependent surface tension is particularly
simple in the context of constant curvature surfaces like cylindrical
and spherical wall/fluid interfaces. In the last decade, different
studies about HS under these geometrical constraints were dedicated
to evaluate the bending rigidity, $k$. On the framework of fundamental
measure theory of Rosenfeld (FMT), and based on the numerical analysis
of the free energy it were found bending rigidity values compatible
with $k=0$ \cite{Bryk_2003a,Konig_2004,Konig_2005}. On the other
hand, on the framework of a similar FMT it was derived an integral
expression for $k$ providing $k\neq0$ and suggesting that Eq. (\ref{eq:gamHadwiger})
is not complete for this system. A second evidence supporting $k\neq0$
result from the re-examination of molecular dynamics (MD) results
\cite{Laird_2012,Blokhuis_2013}. Unfortunately, the smallness of
the obtained maximum value for $k$ at moderate low packing fraction
$\eta$ (smaller than $0.3$) $k\left(\eta=0.2\right)=0.000742$ using
FMT and $k\left(\eta=0.25\right)=0.001\pm0.0008$ using MD, makes
necessary to take them with caution (here $k$ is in $k_{\textrm{B}}T$
units, being $T$ the temperature and $k_{\textrm{B}}$ Boltzmann
constant). On one hand, because the method used to extract the curvature
terms from MD depends on a fit which is very sensible on the adopted
procedure. On the other hand, because FMT is an approximate theory
that produce reliable results but non-necessarily to this higher degree
of accuracy.

In this work I study the higher order curvature dependence of the
surface tension for the HS fluid confined by a spherical and cylindrical
hard walls. The curvature-thermodynamic properties are obtained as
power series in the activity and packing fraction and its coefficients
to order two (for $k$ and $\bar{k}$) or to order three (for $\delta\gamma$,
$2k+\bar{k}$, and the terms of order $R^{-3}$ in Eq. (\ref{eq:gamJK})$ $)
are obtained exactly. An important finding presented here is that
$k\neq0$ on exact grounds and thus Eq. (\ref{eq:gamHadwiger}) is
an approximated expression for the studied system. The effect that
the change of the reference region produces on the system properties
is revisited. The low order series expansions of curvature-thermodynamic
properties are also found by adopting a dividing surface typical of
the scaled particle theory. Using the known terms of the series expansions
complemented with the fitting of available data simple and accurate
analytic expressions are presented for $\delta\gamma$, $2k+\bar{k}$
and for the first time for $k$ and $\bar{k}$. These expressions
transformed to different reference regions are verified by comparing
with available data and it is concluded that they are the most accurate
description for the analyzed properties developed at present.

\section{Expansion in the activity\label{sec:ActivityExp}}

The grand potential of a fluid of point-particles confined in a region
$\mathcal{A}$ by a hard wall (not necessarily a planar wall) is
\begin{equation}
\beta\Omega=-\sum_{i\geq1}\frac{\tau_{i}}{i!}z^{i}\:,\label{eq:Omg1}
\end{equation}
with $\tau_{i}$ the $i$-particles cluster integral of the fluid-in-$\mathcal{A}$
system, $\beta=1/k_{\textrm{B}}T$ its inverse temperature and $z=\Lambda^{-3}\exp\left(-\beta\mu\right)$
its activity. Other magnitudes are: $\mu$ the chemical potential,
$\Lambda$ the de Broglie thermal length and $\beta\Omega=-\ln\Xi$,
being $\Xi$ the grand canonical partition function. For both, spherical
and cylindrical hard walls with a large enough radii $R$, the cluster
integrals can be written as 
\begin{equation}
\frac{\tau_{i}}{i!}=b_{i}V-a_{i}A+c_{i,1}\frac{A}{R}+c_{i,2}\frac{A}{R^{2}}+c_{i,3}\frac{A}{R^{3}}+\ldots\:,\label{eq:Taui1}
\end{equation}
where $V$ is the volume of the system, $A$ is its surface area and
$R$ its radius. At constant temperature the coefficients $b_{i}$,
$a_{i}$ and $c_{i,j}$ with $j\geq1$, are constant. Coefficients
$b_{i}$ and $a_{i}$ are universal and correspond to the bulk cluster
integrals \cite{Hill1956} and to the planar-wall surface cluster
integrals \cite{Bellemans_1962,Sokolowski_1979}, respectively. On
the other hand, $c_{i,j}$ may depend on the shape of the region where
the fluid is confined. Naturally, if the wall is planar $c_{i,j}=0$
for all $j$. Note that $V$, $A$, $R$, $b_{i}$, $a_{i}$ and $c_{i,j}$
refer to a given reference region $\mathcal{B}$ that follows the
spherical/cylindrical symmetry of the one-body distribution function.
Once $\mathcal{B}$ is fixed (through its radius $R=R\left(\mathcal{B}\right)$)
the description given in Eq. (\ref{eq:Taui1}) is unique. Eqs. (\ref{eq:Omg1},
\ref{eq:Taui1}) suggest that
\begin{equation}
\Omega=-PV+\gamma A+C_{1}\frac{A}{R}+C_{2}\frac{A}{R^{2}}+C_{3}\frac{A}{R^{3}}+\ldots\:,\label{eq:Omg2}
\end{equation}
or equivalently
\begin{eqnarray}
\Omega & = & -PV+\gamma\left(R\right)A\:,\label{eq:Omg3}\\
\gamma\left(R\right) & = & \gamma+C_{1}/R+C_{2}/R^{2}+C_{3}/R^{3}+\ldots\:,\label{eq:gamR}
\end{eqnarray}
with $P=\sum_{i\geq1}b_{i}z^{i}$ the pressure of the bulk system
(at the same $z$ and $T$) and
\begin{equation}
\gamma=\sum_{i\geq1}a_{i}z^{i}\,,\: C_{j}=-\sum_{i\geq1}c_{i,j}z^{i}\:,\label{eq:Pgam}
\end{equation}
\begin{equation}
\gamma\left(R\right)=\sum_{i\geq1}\left(a_{i}-\sum_{j\geq1}c_{i,j}R^{-j}\right)z^{i}\:.\label{eq:gamRz}
\end{equation}
Here, $\gamma\left(R\right)$ is the curvature dependent fluid-wall
surface tension and the functions $C_{j}$ are thermodynamic curvature
coefficients. The fluid-wall hard potential induces the formation
of a surface with radius $R_{\textrm{d}}$ where the one body density
distribution, $\rho\left(\mathbf{r}\right)$, drops discontinuously
to zero. The pressure on this zero-density surface, $P_{\textrm{o}}$,
is given by the ideal-gas-like relation $P_{\textrm{o}}/k_{\textrm{B}}T=\rho\left(R_{\textrm{d}}\right)$,
which is known as the wall or contact theorem. Given that $P_{\textrm{o}}=-\frac{dR}{dV}\frac{\partial}{\partial R}\Omega$
one found 
\begin{eqnarray}
P_{\textrm{o}} & = & P+\gamma\frac{2}{R}+C_{1}\frac{1}{R^{2}}-C_{3}\frac{1}{R^{4}}+\ldots\:,\:(sph)\label{eq:Pwsph}\\
P_{\textrm{o}} & = & P+\gamma\frac{1}{R}-C_{2}\frac{1}{R^{3}}-C_{3}\frac{2}{R^{4}}+\ldots\:,\:(cyl)\label{eq:Pwcyl}
\end{eqnarray}
where $P=-\partial\Omega/\partial V$, $\gamma=\partial\Omega/\partial A$,
$C_{j}=\partial\Omega/\partial\left(AR^{-j}\right)$ and it was assumed
that $R-R_{\textrm{d}}$ is a constant length. The mean number of
particles $N=-z\frac{\partial}{\partial z}\,\frac{\Omega}{kT}=\sum_{i\geq1}i\frac{\tau_{i}}{i!}z^{i}$
can also be decomposed as Eqs. (\ref{eq:Omg2}-\ref{eq:gamRz}) resulting
expressions like $N=\rho V+\Gamma A+\Gamma^{(1)}A/R+...$ and $N=\rho V+\Gamma\left(R\right)A$
where the density of the bulk system at the same $z$ and $T$ is
\begin{equation}
\rho=\sum_{i\geq1}ib_{i}z^{i}\:,\label{eq:rho}
\end{equation}
$\Gamma$ is the adsorption on a planar wall, $\Gamma^{(1)}$ is the
first pure curvature adsorption, etc. In fact, the same method apply
to higher order derivatives too, for example to the fluctuation in
the number of particles $\left\langle N^{2}\right\rangle -N^{2}=z\frac{\partial N}{\partial z}=\sum_{i\geq1}i^{2}\frac{\tau_{i}}{i!}z^{i}$.

The Helfrich expansion given in Eq. (\ref{eq:gamJK}) was originally
derived for closed vesicles \cite{Helfrich_1973}. Vesicles are symmetric
in the sense that substance inside and outside are the same, therefore
in this case the sign of $J$ is fixed and can be chosen by convention.
To apply Eq. (\ref{eq:gamJK}) to the unsymmetrical system of a fluid
in contact with a constant curvature hard wall I adopted the usual
convention taking the curvature radius $R_{l}$ as positive for the
fluid outside of the spherical or cylindrical body. Therefore, in
this work Eq. (\ref{eq:gamJK}) reduces to 
\begin{eqnarray}
\gamma\left(R\right) & = & \gamma-2\frac{\delta\gamma}{R}+\frac{2k+\bar{k}}{R^{2}}+\ldots\:,\:(sph)\label{eq:gamsph}\\
\gamma\left(R\right) & = & \gamma-\frac{\delta\gamma}{R}+\frac{k}{2R^{2}}+\ldots\:,\:(cyl)\label{eq:gamcyl}
\end{eqnarray}
where higher order terms like $C_{3(sph)}$ and $C_{3(cyl)}$ are
included. These relations can be compared with Eq. (\ref{eq:gamR})
to obtain 
\begin{eqnarray}
\delta\gamma_{(sph)}=-\frac{1}{2}C_{1(sph)} & \:,\; & ck_{(sph)}=C_{2(sph)}\:,\nonumber \\
\delta\gamma_{(cyl)}=-C_{1(cyl)} & \:,\; & k_{(cyl)}=2C_{2(cyl)}\:,\label{eq:CurvTherm}
\end{eqnarray}
where it was introduced the combined rigidity in short notation $ck=2k+\bar{k}$.
Based on Helfrich's expression, none of the magnitudes $\gamma$,
$\delta$, $k$ and $\bar{k}$ depend on the geometry and thus $(sph)$
and $(cyl)$ labels should be unnecessary.

\section{The HS fluid\label{sec:The-HS-fluid}}

For the fluid of HS with hard repulsion distance $\sigma$ (from
here on when becomes convenient $\sigma$ will be considered $\sigma=1$
for simplicity) in contact with a hard sphere wall $\tau_{2}$ and
$\tau_{3}$ were evaluated in Refs. \cite{Urrutia_2008,Urrutia_2011_b}.
In addition, for the case of a hard cylindrical wall the expression
of $\tau_{2}$ was found in Ref. \cite{Urrutia_2010b}. Instead, the
cluster integrals $\tau_{i}$ for $i>3$ are only partially known
from the values of $b_{i}$ and $a_{i}$ %
\footnote{The coefficient $b_{4}$ is known exactly being its value $-\pi^{2}\left(876\sqrt{2}+94243\pi+8262\,\textrm{ArcCsc}3\right)/90720$
while $b_{5}$ is evaluated using data of Ref. \cite{Clisby_2006}.
Both, $a_{4}$ and $a_{5}$ are evaluated using $a_{2}$, $a_{3}$
and data taken from Ref. \cite{Yang_2013}.\label{fn:footbya}%
}\cite{Clisby_2006}. 
\begin{table}
\begin{centering}
\begin{tabular}{|c|c|c|c|c|}
\hline 
$i$ & 2 & 3 & 4\ref{fn:footbya} & 5\ref{fn:footbya}\tabularnewline
\hline 
\hline 
$b_{i}$ & $-2\pi/3$ & $3\pi^{2}/4$ & $-32.6506\ldots$$ $ & $162.9498822\,(5)$\tabularnewline
\hline 
$a_{i}$ & $-\pi/8$ & $\frac{137}{560}\pi^{2}$ & $-14.3871\,(6)$ & $88.053\,(10)$ \tabularnewline
\hline 
\end{tabular}
\par\end{centering}

\caption{Bulk and surface (planar-wall) coefficients of the cluster integral
$\tau_{i}$ up to $i=5$. $b_{i}$ and $a_{i}$ are expressed in units
of $\sigma^{3}$ and $\sigma^{4}$, respectively.\label{tab:bya} }
\end{table}
The coefficients $b_{i}$ and $a_{i}$ for $i=2,3,4,5$ are listed
in Table \ref{tab:bya}. 
\begin{table}
\begin{centering}
\begin{tabular}{|c|c|c|c|}
\hline 
$i$ & 2 (sph) & 3 (sph) & 2 (cyl)\tabularnewline
\hline 
\hline 
$c_{i,1}$ & $0$ & $-\frac{\pi}{768}\left(9\sqrt{3}+16\pi\right)$ & $ $$0$\tabularnewline
\hline 
$c_{i,2}$ & $-\frac{\pi}{144}$ & $\frac{781}{36288}\pi^{2}$ & $-\frac{\pi}{384}$\tabularnewline
\hline 
$c_{i,3}$ & $0$ & $-\frac{\pi}{1280\sqrt{3}}$ & $0$\tabularnewline
\hline 
\end{tabular}
\par\end{centering}

\caption{Curvature coefficients of the cluster integral $\tau_{2}$ and $\tau_{3}$.
Known terms for spherical and cylindrical hard wall up to $j=3$.
$c_{i,j}$ has units of $\sigma^{4+j}$.\label{tab:coeffc}}
\end{table}
Table \ref{tab:coeffc} summarizes the known curvature coefficients
of $\tau_{i}$ for the spherical and cylindrical hard-walls up to
$j=3$. These coefficients correspond to a choice of the reference
region (RR) $\mathcal{B}$ that coincides with the available region
for the center of each HS, $\mathcal{A}$, with $\rho\left(\mathbf{r}\notin\mathcal{A}\right)=0$.
Thus, under this density-based RR (d-RR) results $R=R_{\textrm{d}}$,
which determines the position of the surface of tension, being $A=A_{\textrm{d}}=4\pi R_{\textrm{d}}$
and $V=V_{\textrm{d}}=4\pi R_{\textrm{d}}/3$. $\textrm{d}$-RR makes
easy the evaluation of $\tau_{i}$ (eg. in d-RR $\tau_{1}=V$). Although
using Eqs. (\ref{eq:Pgam}, \ref{eq:CurvTherm}) it is possible to
explicitly write the curvature-thermodynamic properties as series
expansion in powers of $z$, it is customary to show the results as
function of the packing fraction $\eta=\rho\sigma^{3}\pi/6$. Therefore,
it is necessary to invert the series of Eq. (\ref{eq:rho}) to find
the series of $z\left(\eta\right)$, and then compose series to obtain
the series expansion of each property in powers of $\eta$. Pressure
and surface tension series reproduce the well known virial series
results, eg. $\beta\gamma=-\frac{9\eta^{2}}{2\pi}\left(1+\frac{149}{35}\eta\right)+O\left(\eta^{3}\right)$.
For the spherical wall the exact series expansions of the curvature-thermodynamic
properties gives
\begin{equation}
\beta\,\delta\gamma_{(sph)}=-\frac{9}{64\pi^{2}}\left(9\sqrt{3}+16\pi\right)\eta^{3}+O\left(\eta^{4}\right)\:,\label{eq:dltgamsph}
\end{equation}
\begin{equation}
\beta\, ck_{(sph)}=\frac{\eta^{2}}{4\pi}-\frac{109\eta^{3}}{168\pi}+O\left(\eta^{4}\right)\:,\label{eq:kpkbsph}
\end{equation}
and $\beta C_{3(sph)}=\frac{9\sqrt{3}}{160\pi^{2}}\eta^{3}+O\left(\eta^{4}\right)$.
On the other hand, for the cylindrical wall the exact series expansions
are
\begin{equation}
\beta\,\delta\gamma_{(cyl)}=O\left(\eta^{3}\right)\:,\quad\beta k_{(cyl)}=\frac{3\eta^{2}}{16\pi}+O\left(\eta^{3}\right)\:,\label{eq:dltgamcyl}
\end{equation}
and $\beta C_{3(cyl)}=O\left(\eta^{3}\right)$. Note that $\delta\gamma_{(sph)}$
and $\delta\gamma_{(cyl)}$ are consistent with the Helfrich's expansion
at least up to the highest order for which both are known, i.e. $\delta\gamma_{(sph)}=\delta\gamma_{(cyl)}=\delta\gamma$
up to $O\left(\eta^{3}\right)$. Now, following Helfrich's expansion
one found 
\begin{equation}
\beta\,\delta\gamma=-\frac{9}{64\pi^{2}}\left(9\sqrt{3}+16\pi\right)\eta^{3}+O\left(\eta^{4}\right)\:,\quad(\textrm{d-RR})\label{eq:dltgam0}
\end{equation}

\begin{equation}
\beta\, ck=\frac{\eta^{2}}{4\pi}-\frac{109\eta^{3}}{168\pi}+O\left(\eta^{4}\right)\:,\quad(\textrm{d-RR})\label{eq:kpkb0}
\end{equation}
while the rigidity coefficients are 
\begin{equation}
\beta k=\frac{3\eta^{2}}{16\pi}+O\left(\eta^{3}\right)\:,\;\beta\bar{k}=-\frac{\eta^{2}}{8\pi}+O\left(\eta^{3}\right)\:.\;(\textrm{d-RR})\label{eq:k0}
\end{equation}
Moreover, $c_{i,1(cyl)}=c_{i,1(sph)}/2$ showing that the unknown
coefficient $c_{3,1(cyl)}$ (see Tab. \ref{tab:coeffc}) is indeed
$c_{3,1(cyl)}=-\frac{\pi}{384}\left(9\sqrt{3}+16\pi\right)$.

\section{Different reference regions\label{sec:DiffReferRegn}}

There are at least two different RR adopted in the literature. In
the context of FMT it is usual refer to the d-RR with radius $R=R_{\textrm{d}}$
already discussed in Sec. \ref{sec:The-HS-fluid} \cite{Blokhuis_2013}.
On the other hand, in the scaled particle theory the focus is usually
in the empty region (e-RR) that has a shifted radius of $R=R_{\textrm{e}}=R_{\textrm{d}}-1/2$
(from here on magnitudes referred to e-RR will be labeled with $\textrm{e}$).
No matter the reference adopted $\tau_{i}$ and $\Omega$ remains
invariant because the system remains unmodified. To discuss the change
of reference it is convenient to write the Eq. (\ref{eq:Taui1}) in
matrix notation as 
\begin{equation}
\tau_{i}/i!\:,=\left(\mathbf{b}_{i}\right)_{r}\mathbf{M}_{r}\:,\label{eq:TauibM}
\end{equation}
where up to $O\left(A\, R^{-4}\right)$ the vector of coefficients
is $\mathbf{b}_{i}=\left(b_{i},-a_{i},c_{i,1},c_{i,2},c_{i,3}\right)$
and the column-matrix of measures is $\mathbf{M}=\left(V,A,AR^{-1},AR^{-2},AR^{-3}\right)$.
In fact, both $\mathbf{b}_{i}$ and $\mathbf{M}$ are relative to
the adopted RR, and thus, I introduced the generic label $r$ to make
it explicit. In Sec. \ref{sec:The-HS-fluid}, the system was described
on the basis of measures $\mathbf{M}_{\textrm{d}}=\left(V_{\textrm{d}},A_{\textrm{d}},A_{\textrm{d}}R_{\textrm{d}}^{-1},A_{\textrm{d}}R_{\textrm{d}}^{-2},A_{\textrm{d}}R_{\textrm{d}}^{-3}\right)$
and the corresponding vector of coefficients $\left(\mathbf{b}_{i}\right)_{\textrm{d}}$.
Measures taken with different RR are related by a linear transformation
while the inverse transformation relates the corresponding coefficients.
Here, the procedure is described for a RR with shifted radius $R=R_{u}=R_{\textrm{d}}-u$
that corresponds to measures $\mathbf{M}_{u}=\left(V_{u},A_{u},A_{u}R_{u}^{-1},A_{u}R_{u}^{-2},A_{u}R_{u}^{-3}\right)$
with the obvious definition for the volume and surface area of the
sphere and the cylinder. In matrix form one found 
\begin{equation}
\tau_{i}/i!=\left(\mathbf{b}_{i}\right)_{\textrm{d}}Y^{-1}Y\mathbf{M}_{\textrm{d}}=\left(\mathbf{b}_{i}\right)_{u}\mathbf{M}_{u}\:,\label{eq:TauiRR}
\end{equation}
where $\mathbf{M}_{u}=Y\,\mathbf{M}_{\textrm{d}}$ and $\left(\mathbf{b}_{i}\right)_{u}=\left(\mathbf{b}_{i}\right)_{\textrm{d}}Y^{-1}$.
To build the matrix $Y$, each measure in the $u$-RR frame is written
as a linear function of the measures in the $\textrm{d}$-RR, for
example, $V_{u}=V_{\textrm{d}}+u\, A_{\textrm{d}}-u^{2}A_{\textrm{d}}R^{-1}+\frac{u^{3}}{3}A_{\textrm{d}}R^{-2}$
. Therefore, $Y_{(sph)}$ and $Y_{(cyl)}$ are given by 
\begin{eqnarray}
\left(\begin{array}{ccccc}
1 & u & -u^{2} & u^{3}/3 & 0\\
0 & 1 & -2u & u^{2} & 0\\
0 & 0 & 1 & -u & 0\\
0 & 0 & 0 & 1 & 0\\
0 & 0 & 0 & 0 & 1
\end{array}\right) & \textrm{and} & \left(\begin{array}{ccccc}
1 & u & -u^{2}/2 & 0 & 0\\
0 & 1 & -u & 0 & 0\\
0 & 0 & 1 & 0 & 0\\
0 & 0 & 0 & 1 & u\\
0 & 0 & 0 & 0 & 1
\end{array}\right)\:\label{eq:Ysphcyl}
\end{eqnarray}
respectively (where terms of order $O\left(A_{\textrm{d}}R^{-4}\right)$
were depreciated). The relation between $\left(\mathbf{b}_{i}\right)_{u}$
and $\left(\mathbf{b}_{i}\right)_{\textrm{d}}$ directly implies the
relation for the extensive-like properties: $\left(-P,\gamma,C_{1},C_{2},C_{3}\right)_{u}=\left(-P,\gamma,C_{1},C_{2},C_{3}\right)_{\textrm{d}}Y^{-1}$.
$ $ In particular, $ $taking $u=1/2$ it is possible to derive the
properties in the e-RR for both, the sphere and the cylinder, cases.
Thus, adopting the Helfrich expansion it is found up to $O\left(\eta^{4}\right)$

\begin{equation}
\beta\,\delta\gamma=-\frac{3\eta}{4\pi}\Bigl[1+\eta+\Bigl(\frac{8}{35}+\frac{27\sqrt{3}}{16\pi}\Bigr)\eta^{2}\Bigr]\:,\quad(\textrm{e-RR})\label{eq:dltgameta3}
\end{equation}

\begin{equation}
\beta\, ck=\frac{\eta}{4\pi}\Bigl[1+\frac{\eta}{2}+\Bigl(\frac{81\sqrt{3}}{16\pi}-\frac{289}{105}\Bigr)\eta^{2}\Bigr]\:,\quad(\textrm{e-RR})\label{eq:kpkbeta3}
\end{equation}
while the rigidity constants up to $O\left(\eta^{3}\right)$ take
the form
\begin{equation}
\beta k=\frac{3\eta^{2}}{16\pi}\:,\quad\beta\bar{k}=\frac{\eta}{4\pi}\left(1-\eta\right)\:.\quad(\textrm{e-RR})\label{eq:keta2}
\end{equation}
Besides, the next terms in higher order of $R^{-1}$ are $\beta C_{3\left(sph\right)}=\frac{9\sqrt{3}\eta^{3}}{160\pi^{2}}+O\left(\eta^{4}\right)$
and $\beta C_{3\left(cyl\right)}=\frac{3\eta^{2}}{64\pi}+O\left(\eta^{3}\right)$.
It is noteworthy that $\beta k$ ($\beta C_{3\left(sph\right)}$)
is independent of the chosen radius of the RR to any order in $\eta$
because the fourth (fifth) column of $Y_{\left(cyl\right)}$ ($Y_{\left(sph\right)}$)
is equal to the respective column of the identity matrix.

\section{dependence of the EOS on $\eta$\label{sec:deponeta}}

Expressions for $\gamma$, $\delta\gamma$ and $ck$ as functions
of $\eta$ in e-RR were obtained previously by Reiss \emph{et al.}
using the scaled particle theory (SPT) \cite{Reiss_1960} and more
recently by Hansen-Goss and Roth \cite{HansenGoos_2006} by combining
an FMT approach known as WBII with Eq. (\ref{eq:gamHadwiger}). Both
sets of expressions were only utilized in e-RR while their accuracy
under the adoption of a different RR was not verified. Yet, here I
present a third set of expressions for the $\eta$-dependence of $\gamma$,
$\delta\gamma$ and $ck$ based on the known firsts terms of their
power series in $\eta$. Each of these three sets of functions, complemented
with the bulk pressure EOS to built $\left(P,\gamma,\delta\gamma,ck\right)$,
is compared with FMT and Molecular Dynamics (MD) results to evaluate
its performance in both d-RR and e-RR. Furthermore, under the same
approach I found for the first time expressions for $k$ and $\bar{k}$
as functions of $\eta$.

In the present proposal the functional dependence on $\eta$ for each
property will be obtained using the known exact low order series terms
and including one fitting parameter. To ensure that the thermodynamic
properties are well described regardless the adopted RR the fitting
is done in the framework of d-RR and after transformed to the e-RR.
In making the transformation between different RRs small inaccuracies
in the equations of state may magnify. Therefore, to transform consistently
one must ensure that accurately describe the pressure and the surface
tension. For $\beta P\left(\eta\right)$ one can adopt the very accurate
Kolafa-Malijevsky low density (KM-low) EOS\cite{Kolafa_2004}, however,
I verified that the use of the Carnahan-Starling (CS) EOS instead
of KM-low introduces very small changes and thus the simple and quasi-exact
CS $\beta P\left(\eta\right)/\rho=\left(1+\eta+\eta^{2}-\eta^{3}\right)/\left(1-\eta\right)^{3}$
is utilized. On the other hand, there is not a sufficiently accurate
EOS for the surface tension at present. Here, following the CS rational
expression for $P$ I propose,
\begin{equation}
\beta\gamma=-\frac{9\eta^{2}}{2\pi}\frac{1+\frac{44}{35}\eta+\frac{1}{38}\eta^{2}-3(1-\eta)\eta^{3}}{\left(1-\eta\right)^{3}}\:,\quad(\textrm{d-RR})\label{eq:gamDeta}
\end{equation}
where the parameter is obtained by fitting and then transformed to
the fraction $\frac{1}{38}$. 
\begin{figure}
\begin{centering}
\includegraphics[height=5cm]{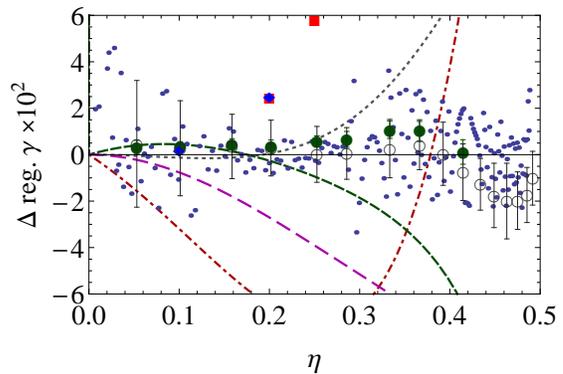}
\par\end{centering}

\caption{(Color online) Difference between several results for the surface
tension and the purposed analytic expression Eq. (\ref{eq:gamDeta})
(in units of $k_{\textrm{B}}T/\sigma^{2}$) by adopting the d-RR.
Plotted is the regularized magnitude $\gamma$ (see the text). Circles
(open and closed) are MD results from Refs. \cite{Laird_2010,Laird_2012},
squares (righted in red and slanted in blue) correspond to FMT results
from Refs. \cite{Bryk_2003a,Blokhuis_2013} while points are MonteCarlo
results. Dot-dashed (red) line is the SPT result, dashed (green)
line is the WBII result, long-dashed (magenta) line corresponds to
the Henderson and Plischke \cite{HendersonD_1987} expression while
dotted (black) line was obtained by Yang \emph{et al}. \cite{Yang_2013}.\label{fig:gam}}
\end{figure}
In Fig. \ref{fig:gam} it is plotted the regularized difference between
the fluid-wall surface tension $\gamma$ taken from different sources
and that given by Eq. (\ref{eq:gamDeta}), where the regularized version
of a magnitude $X$ (reg.$X$) is obtained by dividing $X$ with the
first term of its power series in $\eta$ (eg. $\textrm{reg.}\gamma=\gamma/(-9\eta^{2}/2\pi)$).
The plotted symbols are: open and closed circles for Molecular dynamic
data (MD) taken from Refs. \cite{Laird_2010,Laird_2012}, respectively,
squares (righted in red and slanted in blue) correspond to density
functional FMT results from Refs. \cite{Bryk_2003a,Blokhuis_2013}
and points are MonteCarlo results %
\footnote{MonteCarlo simulation results for $\gamma$ previously published in
Ref. \cite{Yang_2013} were kindly provided by Jung Ho Yang and David
A. Kofke.%
}. It is evident that Eq. (\ref{eq:gamDeta}) accurately describes
the data while other expressions for $\gamma\left(\eta\right)$ (which
are shown as different curves) deviate at $\eta\gtrsim0.25$. For
the curvature-thermodynamic properties I also utilized fitting functions
that combine rational and polynomial forms with one free-parameter.
They are 
\begin{eqnarray}
\beta\delta\gamma & = & -\Bigl(\frac{9}{4\pi}+\frac{81\sqrt{3}}{64\pi^{2}}\Bigr)\,\frac{\eta^{3}\left(1+1.2\eta\right)}{\left(1-\eta\right)^{2}}\:,\quad(\textrm{d-RR})\label{eq:dltDeta}\\
\beta ck & = & \frac{\eta^{2}}{4\pi}\frac{1-\frac{193\eta}{42}}{\left(1-\eta\right)^{2}}-0.54\,\eta^{4}\left(1+3\eta\right)\:,\quad(\textrm{d-RR})\label{eq:2kkbDeta}
\end{eqnarray}
where $1.2$ and $-0.54$ are fitting coefficients.
\begin{figure}
\begin{centering}
\includegraphics[height=8.5cm]{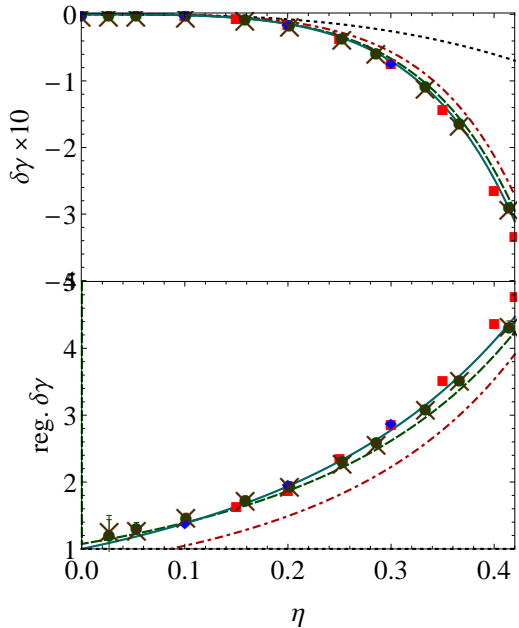}
\par\end{centering}

\caption{(a) Tolman length in d-RR (in units of $k_{\textrm{B}}T/\sigma$).
(b) Regularized Tolman length. Squares are FMT results from Bryk
\emph{et al.} \cite{Bryk_2003a} (rigthed-squares in red) and from
Blokhuis \cite{Blokhuis_2013} (slanted-squares in blue), circles
(green) and crosses are MD results by Laird \emph{et al.} \cite{Laird_2012}.
Continuous line (ligthblue) is the proposed expression, dot-dashed
(red) line is the SPT result, dashed (green) line is the WBII result
while dotted line is the exact series expansion truncated at the last
known term.\label{fig:dltg} }
\end{figure}
\begin{figure}
\begin{centering}
\includegraphics[height=5cm]{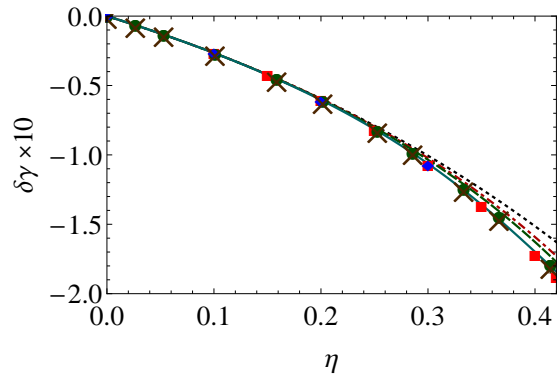}
\par\end{centering}

\caption{Tolman length times the surface tension in e-RR (in units of $k_{\textrm{B}}T/\sigma$).
Lines and symbols were described in Fig. \ref{fig:dltg}.\label{fig:dltgSPT}}
\end{figure}
In Fig. \ref{fig:dltg} (a) the different symbols show the same behavior
with increasing $\eta$, which is well described by WBII and fitted
lines, while SPT curve is slightly above. Reg. $\delta\gamma$ is
presented in Fig. \ref{fig:dltg} (b). There, the FMT data is slightly
separated from MD showing some degree of discrepancy between them,
and also, SPT curve does not reproduce MD and FMT data with sufficient
accuracy. On the other hand, WBII curve correctly describes the data
being the fitted curve the most accurate. Given that the first non-null
coefficient of $\delta\gamma$ as a power series in $\eta$ is wrong
for the SPT formulae in d-RR its failure is not surprising. It is
interesting to compare the FMT results found by Blokhuis and WBII
curve, based on slightly different FMT approaches. At $\eta=0.3$
the difference between both values of $\textrm{reg.}\,\delta\gamma$
is $\sim0.15$. This small difference is relevant because both methods
are accurate in the sense that both involve negligible absolute errors
and thus this disagreement is produced by minimal differences in the
involved approximations. In this and next figures I include results
from a third order polynomial fit in the reciprocal radius of the
MD results \cite{Laird_2012} for the curvature dependent surface
tension of spherical and cylindrical hard walls. These results, plotted
using crosses, were obtained by first writing the $\gamma\left(R\right)$
data in each d-RR and e-RR, and then fitting it. The Fig. \ref{fig:dltgSPT}
shows that Eq. (\ref{eq:dltDeta}) (found in d-RR and then transformed
to e-dd) describes better than the other functional expressions the
behavior of the plotted symbols, and also, that the series in $\eta$
truncated to order $\eta^{3}$ follows the symbols in e-RR much better
than in d-RR. This last advantage of e-RR with respect to d-RR will
be confirmed in the next figures.
\begin{figure}
\begin{centering}
\includegraphics[height=8.5cm]{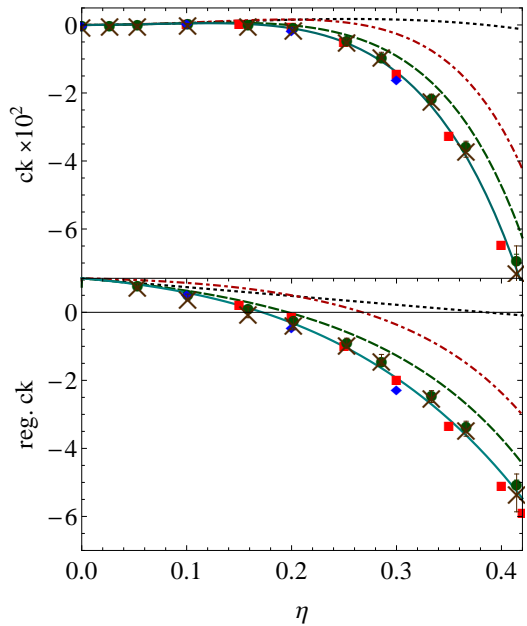}
\par\end{centering}

\caption{(a) Combined curvature rigidity $ck=2k+\bar{k}$ in d-RR (in units
of $k_{\textrm{B}}T$). (b) Magnitude plotted in (a) but in its regularized
form. Lines and symbols were described in Fig. \ref{fig:dltg}.\label{fig:2kkbR}}
\end{figure}
\begin{figure}
\begin{centering}
\includegraphics[height=5cm]{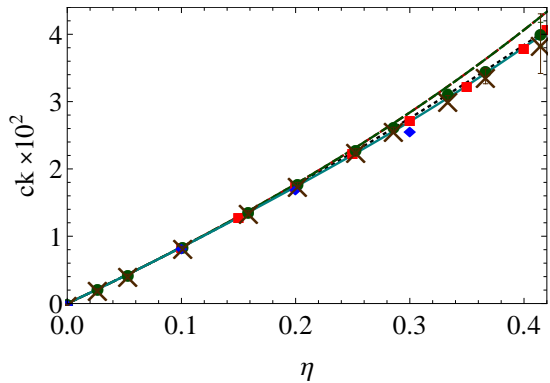}
\par\end{centering}

\caption{Combined curvature rigidity in e-RR (in units of $k_{\textrm{B}}T$).
Lines and symbols were described in Fig. \ref{fig:dltg}.\label{fig:2kkbSPT}}
\end{figure}
 The results for the combined curvature term $ck$ are presented in
Figs. \ref{fig:2kkbR} and \ref{fig:2kkbSPT}. In Fig. \ref{fig:2kkbR}
(a) and (b) is clear that the best description of the behavior of
MD and FMT data is done by Eq. (\ref{eq:2kkbDeta}) being the WBII
curve slightly worse while the worst of the three is that of SPT.
Fig. \ref{fig:2kkbSPT} shows that in e-RR the SPT and WBII produce
nearly identical results which deviate from the symbols at large $\eta$.
Instead, the order-three series truncation given by Eq. (\ref{eq:kpkbeta3})
accurately describes the symbols. Again, one verify that in e-RR the
best curve is provided by the present proposal. The difference between
Blokhuis FMT results and WBII curve in Figs. \ref{fig:2kkbR} (a),
(b) and \ref{fig:2kkbSPT} are apparent and grows with increasing
packing fraction. On general grounds, WBII describes the behavior
of $\delta\gamma$ and $ck$ in both d-RR and e-RR better than SPT
while the proposed expressions make it still better. In addition,
the observed difference between FMT data and WBII suggest that one
of them or both FMT-based results might be turning non-confidence.

Expressions for the bending rigidity $k$ and the Gaussian-curvature
rigidity $\bar{k}$ were, up to my best knowledge, never presented
in the literature. Now, for the bending rigidity the simple dependence
\begin{equation}
\beta k=\frac{3\eta^{2}}{16\pi}\left(1-3\eta\right)\:,\quad(\textrm{e+d-RR})\label{eq:kDeta}
\end{equation}
is proposed based on Eqs. (\ref{eq:k0}, \ref{eq:keta2}) and the
overall analysis of the data of regularized $k$. There the adjusted
coefficient is $-3$.
\begin{figure}
\begin{centering}
\includegraphics[height=8.5cm]{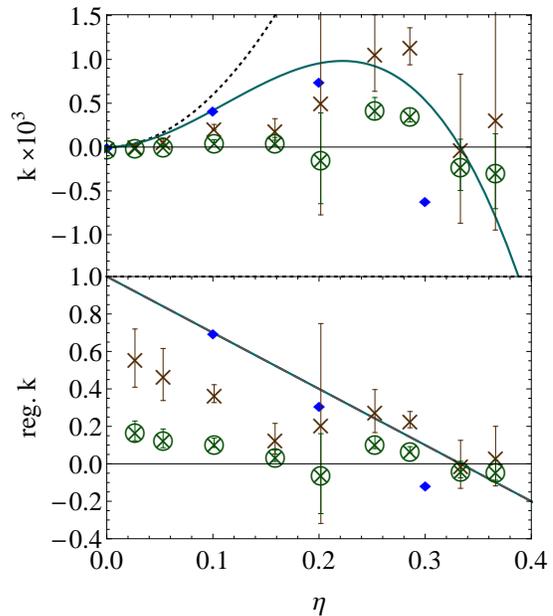}
\par\end{centering}

\caption{(a) Bending constant $k$ for both d-RR and e-RR (in units of $k_{\textrm{B}}T$).
(b) Magnitude plotted in (a) but in its regularized form. Squares
(blue) are FMT results from Blokhuis\cite{Blokhuis_2013}, crosses
and circles with crosses are obtained from MD results by Laird \emph{et
al.}\cite{Laird_2012} Continuous line (lightblue) is the proposed
expression while dotted line is the exact series expansion truncated
at the last known term.\label{fig:k}}
\end{figure}
 The Fig. \ref{fig:k} shows results for $k$ in both d-RR and e-RR,
which should be identical. In Fig. \ref{fig:k} (a) the scale in the
ordinate axes shows that $k$ is a very small magnitude. There, one
can observe a large spreading of the data. Moreover, the MD results
show some degree of inconsistency in the values obtained adopting
the d-RR (crosses) and the e-RR (circles with crosses) while largest
error bars covers the complete range of variation of $k$ with $\eta$.
All these features might indicate that the MD+fit procedure used to
extract the $k$ values is not completely confidence. In addition,
the three FMT values suggest a nearly cubical behavior with a root
at $\eta\sim0.28$. The FMT approach is free from fitting uncertainties
and has a high degree of self-consistence that enables to estimate
absolute errors that are very smalls, eg. $\Delta k\simeq10^{-8}$
for $\eta=0.1$. However, a critical revision of the FMT adopted in
Ref. \cite{Blokhuis_2013} signals that the values for $k$ may be
biased by the inaccuracy of the FMT itself at the high degree of detail
shown in Fig. \ref{fig:k} (a) where the diameter of the circles is
$2\times10^{-4}$ (in $k_{B}T$ units). This makes unclear the confidence
that one should assign to these results to describe by themselves
the subtle behavior of $k\left(\eta\right)$ for the \emph{true} HS
system. In particular, one source of inaccuracy in the adopted FMT
is the use of the Percus-Yevick pressure EOS \cite{Wertheim_1963}
that fails with increasing $\eta$. Based on Fig. \ref{fig:k} (b)
one note that FMT results suggest an overall linear behavior for the
regularized form of $k$. Turning to MD results, at low density it
is clear that none of both sets of data (that correspond to d-RR and
e-RR) point to the correct limiting value reg. $k$$\rightarrow1$
with $\eta\rightarrow0$, and also, that the degree of inconsistency
between both make them non-confidence at low $\eta$. On the other
hand, for $\eta\gtrsim0.22$ the decrease of $k\left(\eta\right)$
and its nearly zero value is well established by the approximate coincidence
of MD and FMT results. Thus, I do not consider the MD results for
$\eta\lesssim0.22$, but I assume a simple linear behavior of regularized
$k$ and make a crude estimate of the slope by considering the FMT
value at the smaller density ($\eta=0.1$) and the MD results for
$\eta\gtrsim0.22$. Based on this analysis, I found a slope of $-3$
which corresponds to Eq. (\ref{eq:kDeta}) and is plotted in the figure.
For the Gaussian-curvature rigidity I propose
\begin{equation}
\beta\bar{k}=-\frac{\eta^{2}}{8\pi}\frac{1-4.2\left(1-6\eta\right)\eta}{\left(1-\eta\right)^{2}}\:,\quad(\textrm{d-RR})\label{eq:kbDeta}
\end{equation}
where $-4.2$ was found by fitting.
\begin{figure}
\begin{centering}
\includegraphics[height=8.5cm]{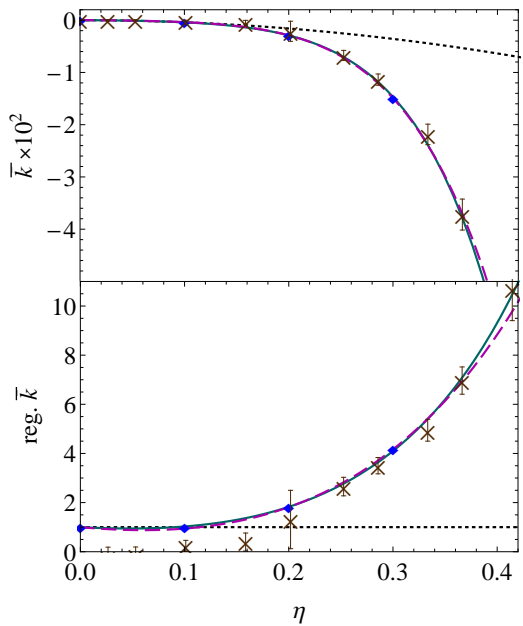}
\par\end{centering}

\caption{(a) Bending constant $\bar{k}$ (in units of $k_{\textrm{B}}T$) in
the framework of d-RR. (b) Magnitude plotted in (a) but in its regularized
form. Lines and symbols were described in Fig. \ref{fig:k}, except
dashed line (magenta) obtained from Eqs. (\ref{eq:2kkbDeta}, \ref{eq:kDeta})
as $ck-2k$.\label{fig:kbR}}
\end{figure}
\begin{figure}
\begin{centering}
\includegraphics[height=5cm]{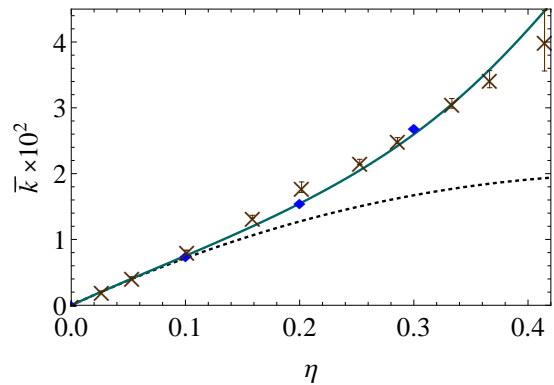}
\par\end{centering}

\caption{Bending constant $\bar{k}$ (in units of $k_{\textrm{B}}T$) in the
framework of e-RR. Lines and symbols were described in Fig. \ref{fig:k}.\label{fig:kbSPT}}
\end{figure}
 In Fig.\ref{fig:kbR} (a) it is plotted $\bar{k}$ in d-RR. There,
one can verify the accurate fitting of Eq. (\ref{eq:kbDeta}) to MD
and FMT data, as well as, the consistence between the fitted $\bar{k}$
and that found from Eqs. (\ref{eq:2kkbDeta}, \ref{eq:kDeta}) as
$ck-2k$. In Fig.\ref{fig:kbR} (b) is apparent that for $\eta\lesssim0.2$
MD data does not has the correct behavior because it does not go to
unit when $\eta\rightarrow0$, this inconsistency is clearly related
to that found in Fig. \ref{fig:k} (b) and makes that MD data for
this range of $\eta$ be non-confidence. Therefore, Eq. (\ref{eq:kbDeta})
is obtained by fitting the three FMT points and those of MD for $\eta\gtrsim0.2$.
The consistence of the approach is verified by the coincidence with
the alternative route to $\bar{k}$ plotted in dashed line. The expression
in Eq. (\ref{eq:kbDeta}) transformed to the e-RR is plotted in Fig.
\ref{fig:kbSPT}. Again, the curve fits very well the data which validates
the obtained description.

\section{Final Remarks}

In this work exact expressions were derived for the Tolman length
and for the combination of curvature terms $2k+\bar{k}$ up to order
three in the packing fraction. Furthermore, the dependence of both
the bending rigidity and the rigidity constant associated with the
Gaussian curvature were found up to order two in the packing fraction.
All these findings are absolute in the sense that they do not imply
the assumption of non-exact EOS and can be readily transformed on
exact grounds to any reference region. Moreover, the functional dependence
of these curvature-thermodynamic properties with the packing fraction
away from $\eta\gtrsim0$ was established on the basis of an approximate
and accurate fitting procedure. 

Based on the truncated power series in the packing fraction, it were
presented definitive evidence showing that for the HS fluid in contact
with hard-curved walls $k\neq0$, and then, that Eq. (\ref{eq:gamHadwiger})
proposed by König \emph{et al.} \cite{Konig_2004,Konig_2005} based
on Hadwiger theorem \cite{Hadwiger1957,Mecke_1998} is not a complete
expression for $\gamma\left(J,K\right)$, at least for the studied
system. The same conclusion extends to the complete morphological
thermodynamic approach which might be considered a good approximate
theory but not an exact one. This result is in good agreement with
that found previously using FMT \cite{Blokhuis_2013}. Even that,
it must be remarked that the use of free energy density functional
theories like FMT for the evaluation of small and sensible quantities
should be done with caution. FMT is an approximate theory for inhomogeneous
fluids, and thus, it is \emph{a priori} unclear which are the boundaries
of reliability of the involved approximations. Particularly, its capability
to describe the subtle behavior of $k$ and its degree of confidence
should be further studied.

Moreover, it is shown that the order $O\left(A\, R^{-4}\right)$ terms
in $\gamma\left(R\right)$ are non-null and thus, that the truncation
up to order $J^{2}$ and $K$ of the Helfrich expansion is also incomplete
and does not enable to accurately describe the known properties of
the HS inhomogeneous system. Notably, the expressions for $\tau_{2(cyl)}\left(R\right)$
and $\tau_{3(sph)}\left(R\right)$, that in this work were truncated
to order $A/R^{3}$, enables to readily extend the results to any
order in powers of $R^{-1}$ showing that the Helfrich expression
is also \emph{approximate} if one truncate it to any finite order
in $R^{-1}$. The accuracy of the analytic expressions for $\gamma$,
$\delta\gamma$, $ck$, $k$ and $\bar{k}$ found by fitting is largely
restricted by the small discrepancies between the different theoretical
methods utilized to obtain their values. In this sense, the development
of a direct MonteCarlo-based method to evaluate the curvature-thermodynamic
properties might be necessary to find more reliable and accurate results.
\begin{acknowledgments}
I am grateful to Jung Ho Yang and David A. Kofke for kindly providing
me with the surface tension MonteCarlo data and to Gabriela Castelletti
and Claudio Pastorino for helpful discussions and comments. This work
was supported by Argentina Grants CONICET PIP-0546/10, UBACyT 20020100200156
and ANPCyT PICT-2011-1887.
\end{acknowledgments}
\bibliographystyle{aipnum4-1}

%

\end{document}